\documentclass[aps,prx,superscriptaddress,amsmath,amssymb,twocolumn,showpacs,floatfix,reprint]{revtex4-2}
\usepackage[colorlinks=true, urlcolor=blue, linkcolor=blue, citecolor=blue, pdftex]{hyperref}
\usepackage[capitalise]{cleveref}
\usepackage{xr}
\usepackage{multirow}
\usepackage[dvipsnames]{xcolor}
\usepackage[utf8]{inputenc}
\usepackage{braket}
\usepackage{graphicx}

\begin{document}
\title{Fermi surface change and $d$-wave superconductivity\\ in the square lattice Kondo-Heisenberg model}

\author{Alexander Nikolaenko}
\thanks{These authors contributed equally. Correspondence should be addressed to rrende@flatironinstitute.org and onikolaienko@g.harvard.edu}
\affiliation{Department of Physics, Harvard University, Cambridge MA 02138, USA}

\author{Riccardo Rende}
\thanks{These authors contributed equally. Correspondence should be addressed to rrende@flatironinstitute.org and onikolaienko@g.harvard.edu}
\affiliation{Center for Computational Quantum Physics, Flatiron Institute, 162 5th Avenue, New York, NY 10010}

\author{Luciano Loris Viteritti}
\affiliation{Institute of Physics, \'{E}cole Polytechnique F\'{e}d\'{e}rale de Lausanne (EPFL), CH-1015 Lausanne, Switzerland}

\author{Subir Sachdev}
\affiliation{Center for Computational Quantum Physics, Flatiron Institute, 162 5th Avenue, New York, NY 10010}
\affiliation{Department of Physics, Harvard University, Cambridge MA 02138, USA}

\author{Ya-Hui Zhang}
\affiliation{Department of Physics and Astronomy, Johns Hopkins University, Baltimore, Maryland 21218, USA}

\date{\today}

\begin{abstract}
We study the two-dimensional Kondo--Heisenberg model on a square lattice, with the conduction electrons away from half-filling, using neural network quantum states. Mapping the ground-state phase diagram as a function of the Kondo and Heisenberg couplings, we identify
({\it i\/}) at weak Kondo coupling, antiferromagnetic N\'eel order with a Fermi surface whose enclosed area counts only the conduction electrons and is insensitive to the N\'eel order, and ({\it ii\/}) at strong coupling, a heavy Fermi liquid with a Fermi surface whose enclosed area counts both the conduction electrons and the spins. In the crossover between these regimes, we find $d_{x^2-y^2}$ superconductivity, evidenced by off-diagonal long-range order in the pair--pair correlations and a pairing-amplitude dome that coexists with the underlying magnetic phase.
Our results establish Fermi volume change and unconventional superconductivity as intrinsic features of the two-dimensional Kondo--Heisenberg model.
\end{abstract}

\maketitle

\begin{figure}[t]
    \begin{center}
\centerline{\includegraphics[width=\columnwidth]{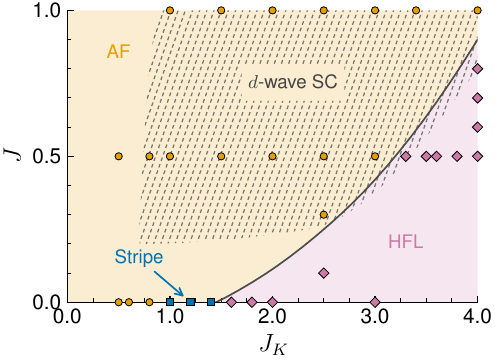}}
        \caption{Ground-state phase diagram of the two-dimensional Kondo-Heisenberg model [see \cref{eq:hamiltonian}] on a square lattice with periodic boundary conditions, shown in the $(J_K,J)$ plane (in units of $t$) at hole doping $\delta=1/4$. We identify an antiferromagnetic N\'eel phase (AF), a stripe magnetic phase (Stripe), a heavy Fermi liquid (HFL), and a $d$-wave superconducting phase ($d$-wave SC) that coexists with the underlying phase. Markers denote the phase assigned to each simulated parameter point.}
        \label{fig:phase_diagram}
    \end{center}
\end{figure}

\emph{Introduction.}
A wide variety of `heavy fermion' intermetallic compounds are well described by Kondo lattice Hamiltonians, which have mobile conducting electrons exchange-coupled to a separate band of stationary spins. These materials have long been of interest for their subtle many-body quantum correlations, leading to large electronic quasiparticle effective masses and unconventional superconductivity. 
Doniach~\cite{doniach1977} argued that the ground state is either an antiferromagnetic metal or a heavy Fermi liquid state, depending on the strength of the Kondo coupling between the localized spins and conducting electrons. Over the years, the theory behind the heavy Fermi liquid state was put on a more rigorous footing, largely owing to large-$N$ approaches~\cite{Read1983,Coleman1987} and Oshikawa's theorem~\cite{Oshikawa2000}. Many attempts were also made both to understand the quantum phase transition between the two competing phases~\cite{Hertz1976,Millis1993,Si01,TSSSMV03,TSMVSS04}, and to investigate the superconducting instabilities in the vicinity of the critical point~\cite{Miyake1986,Scalapino1986,Monod1986}. 

Experimentally, the quantum critical points in various heavy fermion materials were extensively studied by varying pressure and the magnetic field~\cite{Paschen2004,Friedemann2009,Jia2015}, and the enhancement of superconductivity near the critical point was widely recognized~\cite{Hegger2000,Park2006,Pfleiderer2009}. A particularly well studied example is CeCoIn$_5$ \cite{Petrovic2001,Kohori2001,Davis13,Yazdani13,Davis14,Analytis22,park2026}, which has a relatively simple layered square lattice structure and exhibits superconductivity near an interesting quantum phase transition in its normal state.

\begin{figure*}[ht]
    \begin{center}
\centerline{\includegraphics[width=2\columnwidth]{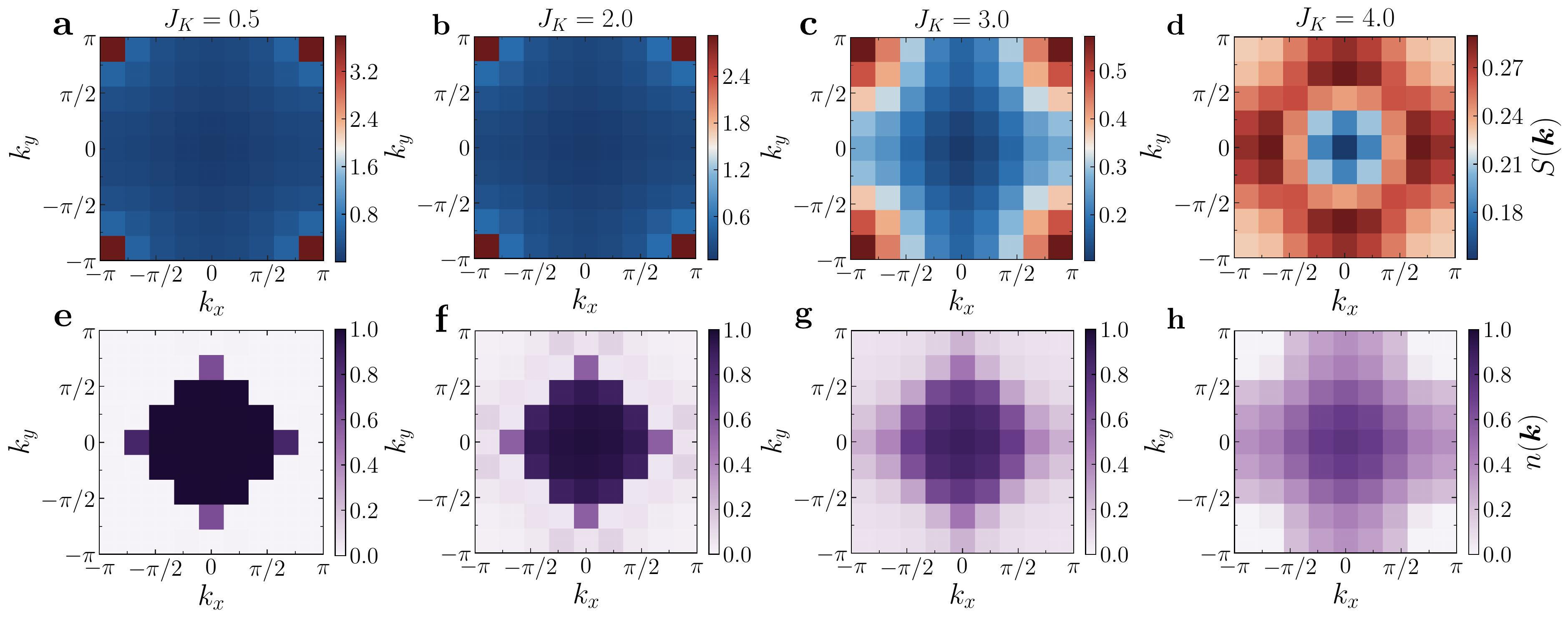}}
    \caption{Static spin structure factor $S(\boldsymbol{k})$ (top row, panels \textbf{a}--\textbf{d}) and conduction-electron momentum distribution $n(\boldsymbol{k})$ (bottom row, panels \textbf{e}--\textbf{h}) in the first Brillouin zone, at fixed $J = 0.5$ and four representative values of the Kondo coupling: $J_K = 0.5$ (panels \textbf{a},\textbf{e}), $J_K = 2.0$ (panels \textbf{b},\textbf{f}), $J_K = 3.0$ (panels \textbf{c},\textbf{g}), and $J_K = 4.0$ (panels \textbf{d},\textbf{h}).}
    \label{fig:Sk_nk}
    \end{center}
\end{figure*}

Despite this strong experimental and theoretical motivation, an unbiased numerical study of the doped two-dimensional model has remained out of reach because each established method runs into a different problem. Auxiliary-field quantum Monte Carlo is free of the sign problem only at half-filling, where particle-hole symmetry protects the simulation; however, the Fermi surface is generically unstable~\cite{Assaad99,Assaad01}; doping the conduction band brings the sign problem back and rules out exactly the regime of interest. The density-matrix renormalization group works well in one dimension~\cite{Sikkema1997,Khait2018,Nikolaenko2024} and on narrow ladders or cylinders~\cite{Dagotto08,Gleis23}, but the entanglement of a truly two-dimensional metal, and in particular of a heavy Fermi liquid with a large Fermi surface, is too large for tensor-network states to handle in practice. Dynamical
mean-field~\cite{Bodensiek2013,Hoshino2014}, variational cluster~\cite{Lenz2017}, and
parton/large-$N$ mean-field~\cite{LiuZhangYu2014,LiuHan2024}
approaches do reach two dimensions and generally point to unconventional ($d$-wave or
related) pairing, but they are all approximate and prone to bias, so none can treat, on an equal footing, the antiferromagnetic, superconducting, and heavy-Fermi-liquid states that compete and coexist in this model.

Neural-network quantum states (NQS)~\cite{carleo2017,rende2026alm,gu2025} get around all of these problems at once. A single, highly expressive ansatz is optimized variationally, free of the sign problem and away from half-filling, on truly periodic two-dimensional square clusters: the antiferromagnetic, heavy-Fermi-liquid, and superconductive states can all be described within the same wave function, and the ground state is picked out by \textit{unbiased} energy minimization rather than assumed~\cite{viteritti2026bias}. The composite local Hilbert space is handled by a transformer architecture designed for such spaces~\cite{rende2026alm}, while a Pfaffian output layer~\cite{viteritti2026bias} encodes magnetic and pairing correlations on the same footing. Our results show a transformation in the electronic Fermi surface, from a `small' Fermi surface involving only the conduction electrons at small Kondo coupling $J_K$, to a `large' Fermi surface involving both the conduction electrons and spin at large $J_K$. Although the small Fermi surface regime also has antiferromagnetic order, this does not significantly modify the Fermi surface because it lies within the antiferromagnetic Brillouin zone boundary. At intermediate $J_K$, our results also show clear evidence of $d$-wave superconductivity. Interestingly, most of the superconducting phase coexists with the antiferromagnetic order.

\emph{Model.}
The Kondo-Heisenberg model is described by:
\begin{equation}
\hat{H} =
-t \sum_{\langle i,j\rangle,\sigma}
\hat{c}^{\dagger}_{i,\sigma}\hat{c}_{j,\sigma}
+ J \sum_{\langle i,j\rangle}
\hat{\boldsymbol{S}}_i\cdot\hat{\boldsymbol{S}}_j
+ J_K \sum_i
\hat{\boldsymbol{S}}_i\cdot\hat{\boldsymbol{s}}_i \ .
\label{eq:hamiltonian}
\end{equation}
The operators $\hat{c}^{\dagger}_{i,\sigma}$ and $\hat{c}_{i,\sigma}$ create and annihilate, respectively, a conduction electron with spin $\sigma=\{\uparrow,\downarrow\}$ at site $i$, while $\hat{\boldsymbol{S}}_i$ denotes the localized spin-$1/2$ operator. The conduction-electron spin is ${\hat{\boldsymbol{s}}_i = \tfrac{1}{2}\sum_{\sigma\sigma'} \hat{c}^{\dagger}_{i,\sigma}\boldsymbol{\tau}_{\sigma\sigma'}\hat{c}_{i,\sigma'}}$, with $\boldsymbol{\tau}$ being the vector of Pauli matrices. The coupling $J$ is the nearest-neighbor superexchange between localized moments, whereas $J_K$ couples each localized moment to the local conduction-electron spin density; the sum $\langle i,j\rangle$ denotes nearest-neighbor pairs. We set $t=1$ and treat $J$ and $J_K$ as independent tuning parameters.

\emph{Results.}
Our main result is the ground-state phase diagram in the $(J_K,J)$ plane,
shown in \cref{fig:phase_diagram}, obtained on the $L=8$ lattice at a fixed
number $N_e$ of conduction electrons, corresponding to a density
$n=N_e/L^2$ and hole doping $\delta=1-n=1/4$. At weak Kondo coupling, the local moments order antiferromagnetically (AF), as expected from the dominant superexchange and RKKY interactions. At strong Kondo coupling, the moments are screened and a heavy Fermi liquid (HFL) emerges. The HFL, while accessible at the mean-field level (see below), is notoriously difficult to stabilize in two dimensions with other variational methods, such as tensor networks, owing to geometric constraints. Here, it is obtained directly from an unbiased variational optimization. In the crossover between these two regimes, we find a $d_{x^2-y^2}$ superconducting region that coexists with the underlying magnetic order. Away from half-filling, superconductivity in this model had previously been suggested only by exact-diagonalization and DMRG studies on quasi-one-dimensional clusters \cite{Dagotto08}; here we establish it on a genuinely two-dimensional lattice. A collinear stripe phase also appears at $J=0$ (see End Matter); we do not investigate its extent in this work. 

We characterize the phases through the local-moment structure factor $S(\boldsymbol{q})=\frac{1}{N}\sum_{ij}e^{i\boldsymbol{q}\cdot(\boldsymbol{R}_i-\boldsymbol{R}_j)}\langle\hat{\boldsymbol{S}}_i\cdot\hat{\boldsymbol{S}}_j\rangle$ and the conduction-electron momentum distribution $n(\boldsymbol{k})=\frac{1}{2}\sum_\sigma\langle\hat{c}^\dagger_{\boldsymbol{k}\sigma}\hat{c}_{\boldsymbol{k}\sigma}\rangle$. Their evolution along the $J=0.5$ cut is shown in \cref{fig:Sk_nk}. At $J_K=0.5$, $S(\boldsymbol{q})$ is sharply peaked at $(\pi,\pi)$, reflecting antiferromagnetic ordering tendencies, and $n(\boldsymbol{k})$ encloses a small Fermi surface of conduction electrons alone. As $J_K$ increases, the $(\pi,\pi)$ peak broadens while the Fermi surface hybridizes with its image shifted by $(\pi,\pi)$. By $J_K\simeq3.0$, the magnetic order is strongly suppressed and the system crosses into the HFL. Crucially, the structure factor is still peaked at $(\pi,\pi)$ and $n(\boldsymbol{k})$ is strongly smeared. At $J_K=4.0$, the moments are fully screened and $n(\boldsymbol{k})$ traces a large Fermi surface whose volume counts both the conduction electrons and the local moments, consistent with Luttinger's theorem~\cite{Oshikawa2000}. The structure factor is peaked at $2k_F$ as a result of the Kohn anomaly~\cite{Kohn1959}.  The pattern of $S(\mathbf q)$ from NQS at $J_K=4.0$ in Fig.~\ref{fig:Sk_nk}(d) can be successfully reproduced by the mean field theory of the HFL (see below).

\begin{figure}[t]
    \begin{center}
\centerline{\includegraphics[width=\columnwidth]{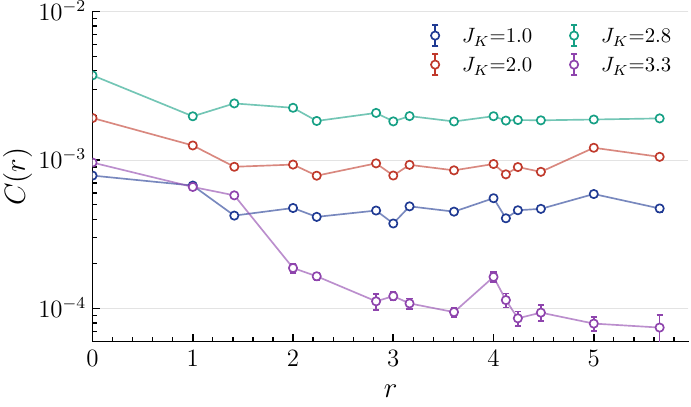}}
    \caption{Connected $d_{x^2-y^2}$ pair--pair correlator $C(\boldsymbol{r})$ as a function of distance $r=|\boldsymbol{r}|$ at $J=0.5$, for $J_K=1.0$, $2.0$, $2.8$, and $3.3$.} \label{fig:superconductivity}
    \end{center}
\end{figure}

To probe pairing, we introduce the $d_{x^2-y^2}$ field $\hat{\Delta}_{\boldsymbol{r}}=\frac14\sum_{\boldsymbol{\eta}} h_{\boldsymbol{r},\boldsymbol{\eta}}\hat{c}_{\boldsymbol{r}\uparrow}\hat{c}_{\boldsymbol{r}+\boldsymbol{\eta}\downarrow}$, with $h_{\boldsymbol{r},\boldsymbol{\eta}}=+1$ ($-1$) on horizontal (vertical) bonds, and measure the connected correlator $C(\boldsymbol{r})=\langle\hat{\Delta}^\dagger_{\boldsymbol{0}}\hat{\Delta}_{\boldsymbol{r}}\rangle-\mathcal{N}_{\boldsymbol{0},\boldsymbol{r}}$, where $\mathcal{N}_{\boldsymbol{0},\boldsymbol{r}}$ removes the disconnected single-particle contribution:
\begin{equation}
\begin{aligned}
\mathcal{N}_{\boldsymbol{0},\boldsymbol{r}} \;=\; \frac{1}{16}\sum_{\boldsymbol{\eta},\boldsymbol{\eta}'} h_{\boldsymbol{0},\boldsymbol{\eta}}\,h_{\boldsymbol{r},\boldsymbol{\eta}'}\,
\Bigl[ &
\langle\hat{c}^{\dagger}_{\boldsymbol{r},\uparrow}\hat{c}_{\boldsymbol{0},\uparrow}\rangle\,
\langle\hat{c}^{\dagger}_{\boldsymbol{r}+\boldsymbol{\eta}',\downarrow}\hat{c}_{\boldsymbol{0}+\boldsymbol{\eta},\downarrow}\rangle \\
&-\;
\langle\hat{c}^{\dagger}_{\boldsymbol{r},\uparrow}\hat{c}_{\boldsymbol{0}+\boldsymbol{\eta},\downarrow}\rangle\,
\langle\hat{c}^{\dagger}_{\boldsymbol{r}+\boldsymbol{\eta}',\downarrow}\hat{c}_{\boldsymbol{0},\uparrow}\rangle
\Bigr].
\end{aligned}
\label{eq:Ndis}
\end{equation}
The long-distance average of $C(\boldsymbol{r})$ defines the order parameter $\Delta=[\frac{1}{\mathcal{M}} \sum_{|\boldsymbol{r}|\ge r_{\max}}C(\boldsymbol{r})]^{1/2}$, where $\mathcal{M}$ is the number of vectors satisfying $|\boldsymbol{r}|\ge r_{\max}$. For $L=8$, we set $r_{\max}=4$. \cref{fig:superconductivity} shows $C(\boldsymbol{r})$ for representative values of $J_K$. For $J_K\simeq1.0$--$3.1$, the correlator saturates to a finite plateau at large distances, the signature of off-diagonal long-range order. In contrast, it decays for $J_K\lesssim1$, deep in the AF phase, and again for $J_K\gtrsim3.1$, once the system has entered the HFL.

\begin{figure}[t]
    \begin{center}
\centerline{\includegraphics[width=\columnwidth]{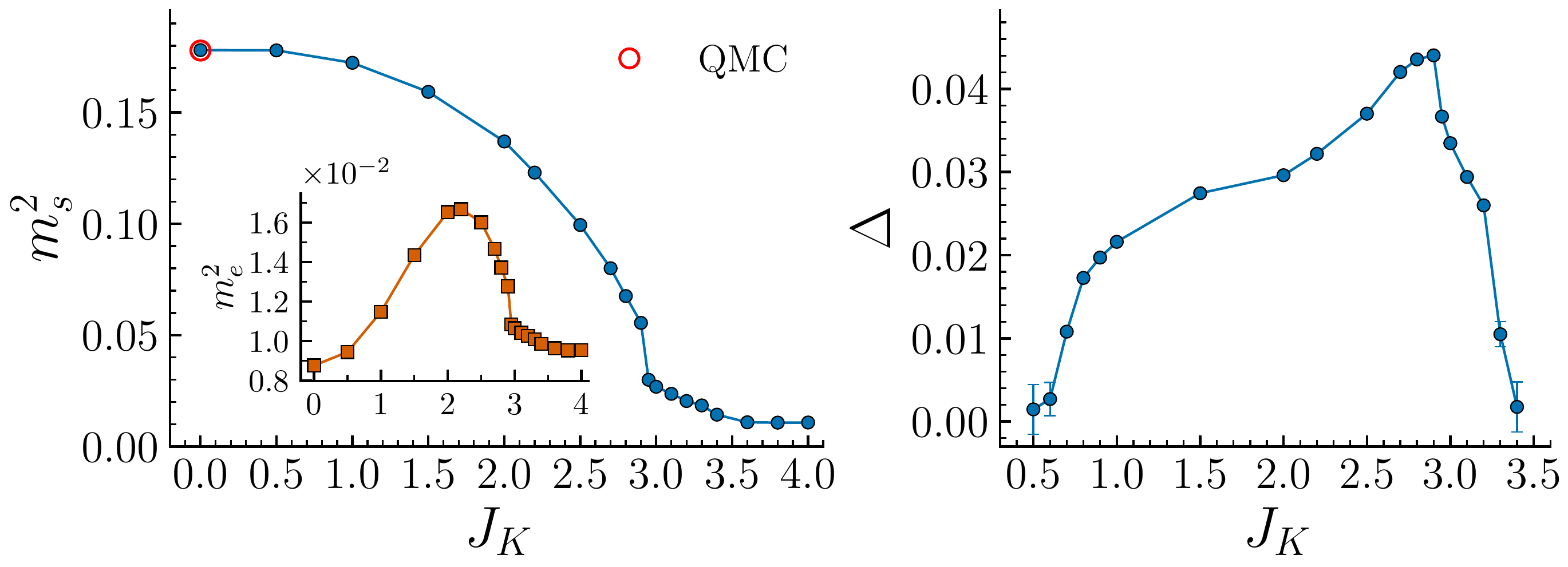}}
    \caption{Left: squared staggered magnetization $m_s^2=S(\pi,\pi)/N$ of the local moments as a function of $J_K$ at $J=0.5$, with QMC data~\cite{sandvik1997prb} at $J_K=0.0$ for comparison. The inset shows the conduction-electron magnetization $m_e^2$. Right: pairing order parameter $\Delta$ as a function of $J_K$ along the same cut.} \label{fig:order_par}
    \end{center}
\end{figure}

\cref{fig:order_par} summarizes the evolution of the two order parameters along the cut $J=0.5$. The squared staggered magnetization $m_s^2=S(\pi,\pi)/N$ (left panel) remains finite up to $J_K\simeq2.8$ and then drops to zero across the transition to the HFL; the inset shows the conduction-electron magnetization $m_e^2$, which is much weaker. The pairing order parameter $\Delta$ (right panel) is essentially zero for $J_K\lesssim0.5$, grows as Kondo screening weakens the antiferromagnetic order, reaches a maximum $\Delta\simeq3.5\times10^{-2}$ at $J_K\simeq2.8$, and vanishes by $J_K\gtrsim3.5$. The resulting dome is peaked between the two non-superconducting phases in the window where magnetic order and Kondo screening compete most strongly.

Together, the long-distance plateau of $C(\boldsymbol{r})$ and the dome shape of $\Delta(J_K)$ provide direct evidence for $d_{x^2-y^2}$ superconductivity in the two-dimensional Kondo--Heisenberg model. The pairing is strongest precisely where antiferromagnetic order is partially suppressed but the moments are not yet fully screened, and its maximum coincides with the collapse of magnetic order at the quantum critical point. This behavior is consistent with the phenomenology of heavy-fermion superconductors and is substantiated by the parton mean-field analysis presented in the Supplementary material.

For future reference and to enable direct comparisons with other variational methods, in \cref{tab:energies} we report the converged variational energy per site $e=\langle\hat{H}\rangle/N$ and the corresponding energy variance per site $\sigma^2 = \left(\langle\hat{H}^2\rangle-\langle\hat{H}\rangle^2\right)/N$ for representative values of the Kondo coupling along the $J=0.5$ cut analyzed in the main text. The variance per site provides an intrinsic, observable-independent measure of the accuracy of the variational state, vanishing for an exact eigenstate.

\emph{Mean-field results.}
We complement the NQS simulations with a mean-field analysis of the Kondo-Heisenberg model. The mean-field treatment reproduces the properties of the antiferromagnetic metal and the heavy Fermi liquid obtained from NQS, providing an independent check on the NQS method. It is less successful for the superconducting state; therefore, in the main text, we restrict the mean-field comparison to the normal state. Given the extensive literature on mean-field approaches to Kondo lattice models (see~\cite{coleman2007} for a historical review and~\cite{Liu2012,park2026} for recent developments), we relegate the technical details to the End Matter and Supplementary Material and present only the main results here. 

\begin{table}[t]
\begin{ruledtabular}
\begin{tabular}{ccc}
$J_K$ & \textbf{Energy} & \textbf{Variance} \\
\hline
$0.5$ & $-1.908883(4)$ & $0.00068(1)$ \\
$2.0$ & $-2.08019(2)$   & $0.012(1)$   \\
$3.0$ & $-2.37980(3)$   & $0.041(1)$   \\
$4.0$ & $-2.84616(2)$   & $0.015(1)$   \\
\end{tabular}
\end{ruledtabular}
\caption{Variational energy per site and energy variance per site for the Kondo-Heisenberg model on the $8\times8$, at hole doping $\delta=1/4$ and $J=0.5$, for the four values of $J_K$ shown in \cref{fig:Sk_nk}. Statistical Monte Carlo errors on the last digit are given in parentheses.}
\label{tab:energies}
\end{table}

\begin{figure}[t]
\begin{minipage}[h]{1\linewidth}
  \center{\includegraphics[width=1\linewidth]{
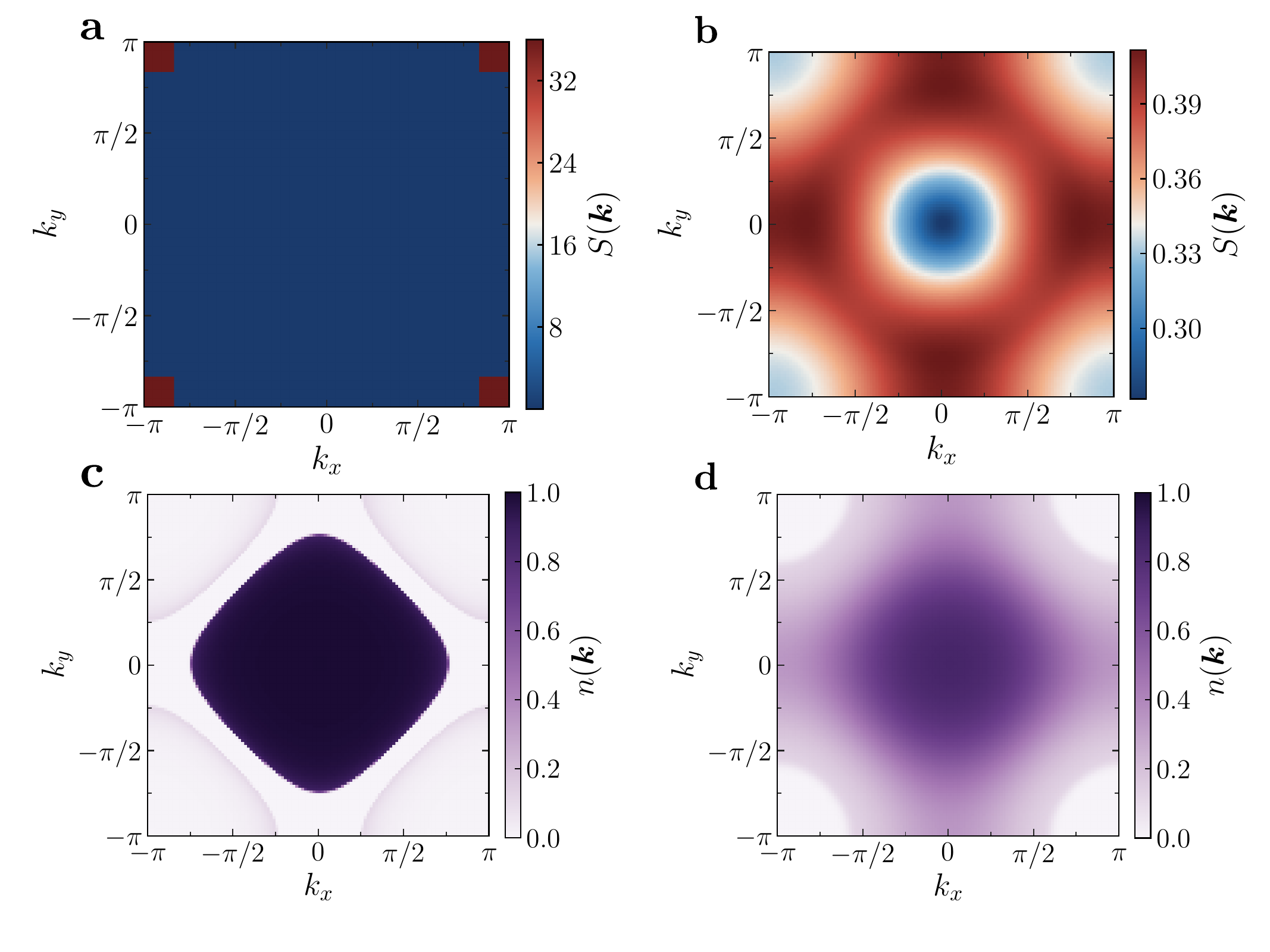}}
\end{minipage}
\caption{Mean field local-moment structure factor $S(\boldsymbol{k})$ (top row) and conduction-electron momentum distribution $n(\boldsymbol{k})$ (bottom row) in the first Brillouin zone, at fixed $J = 0.5$ and three representative values of the Kondo coupling: $J_K = 2.0$ (left, panels $\boldsymbol{a},\boldsymbol{c}$) and $J_K = 4.0$ (right, panels $\boldsymbol{b},\boldsymbol{d}$).}
\label{fig:mean_field_nk}
\end{figure}

Our phase diagram from the mean field theory (see Fig.~\ref{fig:phase_diagram_MF} in the End Matter) is in qualitative agreement with that from the NQS method in Fig.~\ref{fig:phase_diagram}, except that there is a first-order transition between the two metallic phases, which is presumably an artifact of the mean field treatment. At small $J_K$, we also find an antiferromagnetic phase characterized by an ordering vector $Q=(\pi,\pi)$ and staggered order parameters $m_e$ and $m_s$ in the electron and local moment layers. The conduction electron magnetization  $m_e \propto -J_K m_s$ grows linearly with Kondo coupling $J_K$, while local moment magnetization saturates at $m_s=1/2$. The structure factor becomes sharply peaked at the ordering momentum $S(\boldsymbol{k)}=N m_s^2\delta_{\boldsymbol{k},\boldsymbol{Q}}$ (see Fig.~\ref{fig:mean_field_nk} (a)), which is exactly the same as Fig.~\ref{fig:Sk_nk}(a)(b)) from NQS. In Fig.~\ref{fig:mean_field_nk}(c) we show the momentum distribution $n(\mathbf k)$ from the mean field. We can see the electron pocket around the $\Gamma$ point, as in the free fermion level, and its copy at $M=(\pi,\pi)$ connected by the Neel order momentum $\mathbf Q=(\pi,\pi)$. This copied pocket around $\mathcal M$ has smaller spectral weight and is also visible in fig.~\ref{fig:Sk_nk}(f) from NQS, which has less resolution due to the small system size. Actually, the agreement becomes even better if we restrict the mean field also to the system size $L=8$ (see the Supplementary).

At large $J_K$, there is a first-order transition to a heavy Fermi liquid phase with finite hybridization and vanishing magnetic order parameters, $m_s = m_e = 0$. The local moments are absorbed into the Luttinger volume, and the Fermi surface becomes large, with hole pockets centered at $(\pi,\pi)$ [see Fig.~\ref{fig:mean_field_nk}(d)]. The structure factor in Fig.~\ref{fig:mean_field_nk}(b) shows a broad peak at momentum $2k_F$, associated with the Kohn anomaly. Importantly, the mean-field results closely match the NQS predictions (cf. \cref{fig:Sk_nk}). 

\emph{Conclusions.}
We have mapped the ground-state phase diagram of the doped two-dimensional Kondo-Heisenberg model using neural-network quantum states. At weak Kondo coupling, the local moments order antiferromagnetically, and the conduction electrons form a small Fermi surface that is insensitive to the magnetic order. At strong coupling, the moments are screened into a heavy Fermi liquid with a large Fermi surface satisfying Luttinger's theorem. Both phases are 
well described within the mean-field framework, with good qualitative agreement. In the intermediate regime, where magnetic order and Kondo screening compete most strongly, we find robust $d_{x^2-y^2}$ superconductivity, signaled by off-diagonal long-range order in the pair-pair correlations and a pairing dome peaked at the collapse of antiferromagnetic order.

The dome structure, centered near the quantum critical point and vanishing deep in the heavy Fermi liquid phase, together with the $d$-wave symmetry of the superconducting order parameter, indicates that pairing is mediated by the residual short-range antiferromagnetic correlations of the local moments.
This phenomenology closely parallels that observed in layered heavy-fermion compounds such as CeCoIn$_5$~\cite{Petrovic2001,park2026}, and our results establish that it emerges already in the minimal two-dimensional Kondo-Heisenberg model, without invoking band-structure details or additional interactions.
On the methodological side, our results demonstrate that neural-network quantum states can resolve the delicate competition between magnetic, striped, superconducting, and heavy Fermi liquid states in a genuinely two-dimensional Kondo lattice -- a regime that has remained out of reach for sign-problem-free quantum Monte Carlo and is challenging for tensor-network methods. Natural extensions of this work include a systematic finite-size scaling of the pairing order parameter, the doping dependence of the superconducting dome, and the nature of the transition between the antiferromagnet and the heavy Fermi liquid in the presence of frustrating exchange interactions, where the possible existence of deconfined criticality or fractionalized intermediate phases~\cite{Si01,TSSSMV03,TSMVSS04} remains a central open question in heavy-fermion physics.

\begin{acknowledgments}
We thank Antoine Georges and Shiwei Zhang for useful discussions. 
A.N. and S.S. were supported by the U.S. National Science Foundation grant No. DMR-2245246 and by the Simons Collaboration on Ultra-Quantum Matter which is a grant from the Simons Foundation (651440, S.S.). YHZ was supported by the  National Science Foundation under Grant No. DMR-2237031.
The Flatiron Institute is a division of the Simons Foundation. The simulations presented in this work required a total computational budget of approximately $30\,000$ GPU hours on NVIDIA H200 GPUs.
\end{acknowledgments}

\bibliography{refs}

@ARTICLE{Davis14,
       author = {{Van Dyke}, John S. and {Massee}, Freek and {Allan}, Milan P. and {Davis}, J.~C. S{\'e}amus and {Petrovic}, Cedomir and {Morr}, Dirk K.},
        title = "{Direct evidence for a magnetic $f$-electron-mediated pairing mechanism of heavy-fermion superconductivity in CeCoIn$_{5}$}",
      journal = {Proceedings of the National Academy of Science},
         year = 2014,
        month = aug,
       volume = {111},
       number = {32},
        pages = {11663-11667},
          doi = {10.1073/pnas.1409444111},
archivePrefix = {arXiv},
       eprint = {1405.5883},
 primaryClass = {cond-mat.supr-con},
       adsurl = {https://ui.adsabs.harvard.edu/abs/2014PNAS..11111663V}
}

@article{Analytis22,
author = {Nikola Maksimovic  and Daniel H. Eilbott  and Tessa Cookmeyer  and Fanghui Wan  and Jan Rusz  and Vikram Nagarajan  and Shannon C. Haley  and Eran Maniv  and Amanda Gong  and Stefano Faubel  and Ian M. Hayes  and Ali Bangura  and John Singleton  and Johanna C. Palmstrom  and Laurel Winter  and Ross McDonald  and Sooyoung Jang  and Ping Ai  and Yi Lin  and Samuel Ciocys  and Jacob Gobbo  and Yochai Werman  and Peter M. Oppeneer  and Ehud Altman  and Alessandra Lanzara  and James G. Analytis },
title = "{Evidence for a delocalization quantum phase transition without symmetry breaking in CeCoIn$_5$}",
journal = {Science},
volume = {375},
number = {6576},
pages = {76-81},
year = {2022},
doi = {10.1126/science.aaz4566}}

@article{Petrovic2001,
  author    = {C. Petrovic and R. Movshovich and M. Jaime and P. G. Pagliuso and M. F. Hundley and J. L. Sarrao and Z. Fisk and J. D. Thompson},
  title     = "{A new heavy-fermion superconductor CeCoIn$_5$: A relative of the cuprates?}",
  journal   = {J. Phys.: Condens. Matter},
  volume    = {13},
  pages     = {L337--L342},
  year      = {2001},
  doi       = {10.1088/0953-8984/13/17/103}
}

@article{Kohori2001,
  author    = {Y. Kohori and Y. Yamato and Y. Iwamoto and T. Kohara and E. D. Bauer and M. B. Maple and J. L. Sarrao},
  title = "{NMR and NQR studies of the heavy fermion superconductors CeTIn$_5$ (T = Co and Ir)}",
  journal   = {Phys. Rev. B},
  volume    = {64},
  pages     = {134526},
  year      = {2001},
  doi       = {10.1103/PhysRevB.64.134526}
}

@ARTICLE{Davis13,
       author = {{Allan}, M.~P. and {Massee}, F. and {Morr}, D.~K. and {Van Dyke}, J. and {Rost}, A.~W. and {Mackenzie}, A.~P. and {Petrovic}, C. and {Davis}, J.~C.},
        title = "{Imaging Cooper pairing of heavy fermions in CeCoIn$_{5}$}",
      journal = {Nature Physics},
         year = 2013,
        month = aug,
       volume = {9},
       number = {8},
        pages = {468-473},
          doi = {10.1038/nphys2671},
archivePrefix = {arXiv},
       eprint = {1303.4416},
 primaryClass = {cond-mat.supr-con},
       adsurl = {https://ui.adsabs.harvard.edu/abs/2013NatPh...9..468A}
}

@ARTICLE{Yazdani13,
       author = {{Zhou}, Brian B. and {Misra}, Shashank and {da Silva Neto}, Eduardo H. and {Aynajian}, Pegor and {Baumbach}, Ryan E. and {Thompson}, J.~D. and {Bauer}, Eric D. and {Yazdani}, Ali},
        title = "{Visualizing nodal heavy fermion superconductivity in CeCoIn$_{5}$}",
      journal = {Nature Physics},
         year = 2013,
        month = aug,
       volume = {9},
       number = {8},
        pages = {474-479},
          doi = {10.1038/nphys2672},
archivePrefix = {arXiv},
       eprint = {1307.3787},
 primaryClass = {cond-mat.supr-con},
       adsurl = {https://ui.adsabs.harvard.edu/abs/2013NatPh...9..474Z}
}

@ARTICLE{Dagotto08,
       author = {{Xavier}, J.~C. and {Dagotto}, E.},
        title = "{Robust $d$-Wave Pairing Correlations in the Heisenberg Kondo Lattice Model}",
      journal = {Phys. Rev. Lett.},
         year = 2008,
        month = apr,
       volume = {100},
       number = {14},
          eid = {146403},
        pages = {146403},
          doi = {10.1103/PhysRevLett.100.146403},
archivePrefix = {arXiv},
       eprint = {0803.1486},
 primaryClass = {cond-mat.str-el},
       adsurl = {https://ui.adsabs.harvard.edu/abs/2008PhRvL.100n6403X}
}

@article{Kohn1959,
  title = "{Image of the Fermi Surface in the Vibration Spectrum of a Metal}",
  author = {Kohn, W.},
  journal = {Phys. Rev. Lett.},
  volume = {2},
  issue = {9},
  pages = {393--394},
  numpages = {0},
  year = {1959},
  month = {May},
  publisher = {American Physical Society},
  doi = {10.1103/PhysRevLett.2.393},
  url = {https://link.aps.org/doi/10.1103/PhysRevLett.2.393}
}

@ARTICLE{Gleis23,
       author = {{Gleis}, Andreas and {Li}, Jheng-Wei and {von Delft}, Jan},
        title = "{Controlled Bond Expansion for Density Matrix Renormalization Group Ground State Search at Single-Site Costs}",
      journal = {Phys. Rev. Lett.},
         year = 2023,
        month = jun,
       volume = {130},
       number = {24},
          eid = {246402},
        pages = {246402},
          doi = {10.1103/PhysRevLett.130.246402},
archivePrefix = {arXiv},
       eprint = {2207.14712},
 primaryClass = {cond-mat.str-el},
       adsurl = {https://ui.adsabs.harvard.edu/abs/2023PhRvL.130x6402G}
}

@ARTICLE{Si01,
       author = {{Si}, Qimiao and {Rabello}, Silvio and {Ingersent}, Kevin and {Smith}, J. Lleweilun},
        title = "{Locally critical quantum phase transitions in strongly correlated metals}",
      journal = {Nature},
         year = 2001,
        month = oct,
       volume = {413},
       number = {6858},
        pages = {804-808},
          doi = {10.1038/35101507},
archivePrefix = {arXiv},
       eprint = {cond-mat/0011477},
 primaryClass = {cond-mat.str-el},
       adsurl = {https://ui.adsabs.harvard.edu/abs/2001Natur.413..804S}
}

@misc{coleman2007,
      title="{Heavy Fermions: electrons at the edge of magnetism}", 
      author={P. Coleman},
      year={2007},
      eprint={cond-mat/0612006},
      archivePrefix={arXiv},
      primaryClass={cond-mat.str-el}
}

@article{Liu2012,
       author = {{Liu}, Yu and {Li}, Huan and {Zhang}, Guang-Ming and {Yu}, Lu},
        title = "{$d$-wave superconductivity induced by short-range antiferromagnetic correlations in the two-dimensional Kondo lattice model}",
      journal = {Phys. Rev. B},
         year = 2012,
        month = jul,
       volume = {86},
       number = {2},
          eid = {024526},
        pages = {024526},
          doi = {10.1103/PhysRevB.86.024526},
archivePrefix = {arXiv},
       eprint = {1204.1434},
 primaryClass = {cond-mat.str-el},
       adsurl = {https://ui.adsabs.harvard.edu/abs/2012PhRvB..86b4526L}
}

@ARTICLE{Assaad99,
       author = {{Assaad}, F.~F.},
        title = "{Quantum Monte Carlo Simulations of the Half-Filled Two-Dimensional Kondo Lattice Model}",
      journal = {Phys. Rev. Lett.},
         year = 1999,
        month = jul,
       volume = {83},
       number = {4},
        pages = {796-799},
          doi = {10.1103/PhysRevLett.83.796},
archivePrefix = {arXiv},
       eprint = {cond-mat/9904178},
 primaryClass = {cond-mat.str-el},
       adsurl = {https://ui.adsabs.harvard.edu/abs/1999PhRvL..83..796A}
}

@ARTICLE{Assaad01,
       author = {{Capponi}, S. and {Assaad}, F.~F.},
        title = "{Spin and charge dynamics of the ferromagnetic and antiferromagnetic two-dimensional half-filled Kondo lattice model}",
      journal = {Phys. Rev. B},
         year = 2001,
        month = apr,
       volume = {63},
       number = {15},
          eid = {155114},
        pages = {155114},
          doi = {10.1103/PhysRevB.63.155114},
archivePrefix = {arXiv},
       eprint = {cond-mat/0010393},
 primaryClass = {cond-mat.str-el},
       adsurl = {https://ui.adsabs.harvard.edu/abs/2001PhRvB..63o5114C}
}

@misc{park2026,
      title="{Superconductivity and fractionalized magnetic excitations in CeCoIn$_5$}", 
      author={Pyeongjae Park and Shang-Shun Zhang and Pietro M. Bonetti and Andrey A. Podlesnyak and Daniel M. Pajerowski and Matthew B. Stone and C. Petrovic and C. Stock and Subir Sachdev and Cristian D. Batista and Andrew D. Christianson},
      year={2026},
      eprint={2604.02481},
      archivePrefix={arXiv}
}

@article{viteritti2025prb,
  title = "{Transformer wave function for two dimensional frustrated magnets: Emergence of a spin-liquid phase in the Shastry-Sutherland model}",
  author = {Viteritti, Luciano Loris and Rende, Riccardo and Parola, Alberto and Goldt, Sebastian and Becca, Federico},
  journal = {Phys. Rev. B},
  volume = {111},
  issue = {13},
  pages = {134411},
  numpages = {15},
  year = {2025},
  month = {Apr},
  publisher = {American Physical Society},
  doi = {10.1103/PhysRevB.111.134411},
  url = {https://link.aps.org/doi/10.1103/PhysRevB.111.134411}
}

@misc{gu2025,
      title="{Solving the Hubbard model with Neural Quantum States}", 
      author={Yuntian Gu and Wenrui Li and Heng Lin and Bo Zhan and Ruichen Li and Yifei Huang and Di He and Yantao Wu and Tao Xiang and Mingpu Qin and Liwei Wang and Dingshun Lv},
      year={2025},
      eprint={2507.02644},
      archivePrefix={arXiv},
      primaryClass={cond-mat.str-el},
      url={https://arxiv.org/abs/2507.02644}, 
}

@misc{viteritti2026approaching,
      title="{Approaching the Thermodynamic Limit with Neural-Network Quantum States}", 
      author={Luciano Loris Viteritti and Riccardo Rende and Subir Sachdev and Giuseppe Carleo},
      year={2026},
      eprint={2602.02665},
      archivePrefix={arXiv},
      primaryClass={cond-mat.str-el} 
}

@article{viteritti2023prl,
  title = {Transformer Variational Wave Functions for Frustrated Quantum Spin Systems},
  author = {Viteritti, Luciano Loris and Rende, Riccardo and Becca, Federico},
  journal = {Phys. Rev. Lett.},
  volume = {130},
  issue = {23},
  pages = {236401},
  numpages = {6},
  year = {2023},
  month = {Jun},
  publisher = {American Physical Society},
  doi = {10.1103/PhysRevLett.130.236401},
  url = {https://link.aps.org/doi/10.1103/PhysRevLett.130.236401}
}

@article{rende2024prr,
  title = "{Mapping of attention mechanisms to a generalized Potts model}",
  author = {Rende, Riccardo and Gerace, Federica and Laio, Alessandro and Goldt, Sebastian},
  journal = {Phys. Rev. Res.},
  volume = {6},
  issue = {2},
  pages = {023057},
  numpages = {10},
  year = {2024},
  month = {Apr},
  publisher = {American Physical Society},
  doi = {10.1103/PhysRevResearch.6.023057},
  url = {https://link.aps.org/doi/10.1103/PhysRevResearch.6.023057}
}

@article{carleo2017,
   title={Solving the quantum many-body problem with artificial neural networks},
   volume={355},
   ISSN={1095-9203},
   url={http://dx.doi.org/10.1126/science.aag2302},
   DOI={10.1126/science.aag2302},
   number={6325},
   journal={Science},
   publisher={American Association for the Advancement of Science (AAAS)},
   author={Carleo, Giuseppe and Troyer, Matthias},
   year={2017},
   month=Feb, pages={602–606} }

@article{doniach1977,
title = "{The Kondo lattice and weak antiferromagnetism}",
journal = {Physica B+C},
volume = {91},
pages = {231-234},
year = {1977},
issn = {0378-4363},
doi = {https://doi.org/10.1016/0378-4363(77)90190-5},
url = {https://www.sciencedirect.com/science/article/pii/0378436377901905},
author = {S. Doniach},
}

@article{Read1983,
doi = {10.1088/0022-3719/16/17/014},
url = {https://doi.org/10.1088/0022-3719/16/17/014},
year = {1983},
month = {jun},
publisher = {},
volume = {16},
number = {17},
pages = {3273},
author = {N Read and D M Newns},
title = "{On the solution of the Coqblin-Schreiffer Hamiltonian by the large-$N$ expansion technique}",
journal = {Journal of Physics C: Solid State Physics}
}

@article{Coleman1987,
  title = {Mixed valence as an almost broken symmetry},
  author = {Coleman, Piers},
  journal = {Phys. Rev. B},
  volume = {35},
  issue = {10},
  pages = {5072--5116},
  numpages = {0},
  year = {1987},
  month = {Apr},
  publisher = {American Physical Society},
  doi = {10.1103/PhysRevB.35.5072},
  url = {https://link.aps.org/doi/10.1103/PhysRevB.35.5072}
}

@article{Oshikawa2000,
  title = "{Topological Approach to Luttinger's Theorem and the Fermi Surface of a Kondo Lattice}",
  author = {Oshikawa, Masaki},
  journal = {Phys. Rev. Lett.},
  volume = {84},
  issue = {15},
  pages = {3370--3373},
  numpages = {0},
  year = {2000},
  month = {Apr},
  publisher = {American Physical Society},
  doi = {10.1103/PhysRevLett.84.3370},
  url = {https://link.aps.org/doi/10.1103/PhysRevLett.84.3370}
}

@ARTICLE{TSSSMV03,
   author = {{Senthil}, T. and {Sachdev}, S. and {Vojta}, M.},
    title = "{Fractionalized Fermi Liquids}",
  journal = {Phys. Rev. Lett.},
   eprint = {cond-mat/0209144},
     year = 2003,
    month = may,
   volume = 90,
   number = 21,
      eid = {216403},
    pages = {216403},
      doi = {10.1103/PhysRevLett.90.216403},
   adsurl = {http://adsabs.harvard.edu/abs/2003PhRvL..90u6403S}
}

@article{TSMVSS04,
   author = {{Senthil}, T. and {Vojta}, M. and {Sachdev}, S.},
    title = "{Weak magnetism and non-Fermi liquids near heavy-fermion critical points}",
  journal = {Phys. Rev. B},
   eprint = {cond-mat/0305193},
     year = 2004,
    month = jan,
   volume = 69,
   number = 3,
      eid = {035111},
    pages = {035111},
      doi = {10.1103/PhysRevB.69.035111},
   adsurl = {http://adsabs.harvard.edu/abs/2004PhRvB..69c5111S}
}

@misc{rende2026alm,
      title="{Transformer Neural-Network Quantum States for lattice models of spins and fermions: Application to the Ancilla Layer Model}", 
      author={Riccardo Rende and Alexander Nikolaenko and Luciano Loris Viteritti and Subir Sachdev and Ya-Hui Zhang},
      year={2026},
      eprint={2603.02316},
      archivePrefix={arXiv},
      primaryClass={cond-mat.str-el}
}

@article{Sorella1998,
  title = "{Green Function Monte Carlo with Stochastic Reconfiguration}",
  author = {Sorella, Sandro},
  journal = {Phys. Rev. Lett.},
  volume = {80},
  issue = {20},
  pages = {4558--4561},
  numpages = {0},
  year = {1998},
  month = {May},
  publisher = {American Physical Society},
  doi = {10.1103/PhysRevLett.80.4558},
  url = {https://link.aps.org/doi/10.1103/PhysRevLett.80.4558}
}

@article{rende2024stochastic,
   title={A simple linear algebra identity to optimize large-scale neural network quantum states},
   volume={7},
   pages = 260,
   ISSN={2399-3650},
   DOI={10.1038/s42005-024-01732-4},
   number={1},
   journal={Communications Physics},
   publisher={Springer Science and Business Media LLC},
   author={Rende, Riccardo and Viteritti, Luciano Loris and Bardone, Lorenzo and Becca, Federico and Goldt, Sebastian},
   year={2024},
   month=aug }

@article{chen2024empowering,
  title={Empowering deep neural quantum states through efficient optimization},
  author={Chen, Ao and Heyl, Markus},
  journal={Nature Physics},
  volume={20},
  number={9},
  pages={1476--1481},
  year={2024},
  publisher={Nature Publishing Group UK London},
  doi = {10.1038/s41567-024-02566-1}
}

@article{Monod1986,
  title = {Possible superconductivity in nearly antiferromagnetic itinerant fermion systems},
  author = {B\'eal-Monod, M. T. and Bourbonnais, C. and Emery, V. J.},
  journal = {Phys. Rev. B},
  volume = {34},
  issue = {11},
  pages = {7716--7720},
  numpages = {0},
  year = {1986},
  month = {Dec},
  publisher = {American Physical Society},
  doi = {10.1103/PhysRevB.34.7716},
  url = {https://link.aps.org/doi/10.1103/PhysRevB.34.7716}
}

@article{Friedemann2009,
	author = {Friedemann, S. and Westerkamp, T. and Brando, M. and Oeschler, N. and Wirth, S. and Gegenwart, P. and Krellner, C. and Geibel, C. and Steglich, F.},
	journal = {Nature Physics},
	number = {7},
	pages = {465--469},
	title = "{Detaching the antiferromagnetic quantum critical point from the Fermi-surface reconstruction in YbRh$_2$Si$_2$}",
	volume = {5},
	year = {2009}}

@article{Pfleiderer2009,
  title = {Superconducting phases of $f$-electron compounds},
  author = {Pfleiderer, Christian},
  journal = {Rev. Mod. Phys.},
  volume = {81},
  issue = {4},
  pages = {1551--1624},
  numpages = {0},
  year = {2009},
  month = {Nov},
  publisher = {American Physical Society},
  doi = {10.1103/RevModPhys.81.1551},
  url = {https://link.aps.org/doi/10.1103/RevModPhys.81.1551}
}

@article{Sikkema1997,
  title = "{Spin Gap in a Doped Kondo Chain}",
  author = {Sikkema, Arnold E. and Affleck, Ian and White, Steven R.},
  journal = {Phys. Rev. Lett.},
  volume = {79},
  issue = {5},
  pages = {929--932},
  numpages = {0},
  year = {1997},
  month = {Aug},
  publisher = {American Physical Society},
  doi = {10.1103/PhysRevLett.79.929},
  url = {https://link.aps.org/doi/10.1103/PhysRevLett.79.929}
}

@Article{Nikolaenko2024,
	title={{Numerical signatures of ultra-local criticality in a one dimensional Kondo lattice model}},
	author={Alexander Nikolaenko and Ya-Hui Zhang},
	journal={SciPost Phys.},
	volume={17},
	pages={034},
	year={2024},
	publisher={SciPost},
	doi={10.21468/SciPostPhys.17.2.034},
	url={https://scipost.org/10.21468/SciPostPhys.17.2.034},
}

@article{
Khait2018,
author = {Ilia Khait  and Patrick Azaria  and Claudius Hubig  and Ulrich Schollwöck  and Assa Auerbach },
title = "{Doped Kondo chain, a heavy Luttinger liquid}",
journal = {Proceedings of the National Academy of Sciences},
volume = {115},
number = {20},
pages = {5140-5144},
year = {2018},
doi = {10.1073/pnas.1719374115}}

@article{Park2006,
	author = {Park, Tuson and Ronning, F. and Yuan, H. Q. and Salamon, M. B. and Movshovich, R. and Sarrao, J. L. and Thompson, J. D.},
	journal = {Nature},
	number = {7080},
	pages = {65--68},
	title = "{Hidden magnetism and quantum criticality in the heavy fermion superconductor CeRhIn$_5$}",
	volume = {440},
	year = {2006}}

@article{
Jia2015,
author = {Lin Jiao  and Ye Chen  and Yoshimitsu Kohama  and David Graf  and E. D. Bauer  and John Singleton  and Jian-Xin Zhu  and Zongfa Weng  and Guiming Pang  and Tian Shang  and Jinglei Zhang  and Han-Oh Lee  and Tuson Park  and Marcelo Jaime  and J. D. Thompson  and Frank Steglich  and Qimiao Si  and H. Q. Yuan },
title = "{Fermi surface reconstruction and multiple quantum phase transitions in the antiferromagnet CeRhIn$_5$}",
journal = {Proceedings of the National Academy of Sciences},
volume = {112},
number = {3},
pages = {673-678},
year = {2015},
doi = {10.1073/pnas.1413932112}}

@article{Paschen2004,
	author = {Paschen, S. and L{\"u}hmann, T. and Wirth, S. and Gegenwart, P. and Trovarelli, O. and Geibel, C. and Steglich, F. and Coleman, P. and Si, Q.},
	date = {2004/12/01},
	doi = {10.1038/nature03129},
	id = {Paschen2004},
	isbn = {1476-4687},
	journal = {Nature},
	number = {7019},
	pages = {881--885},
	title = {Hall-effect evolution across a heavy-fermion quantum critical point},
	url = {https://doi.org/10.1038/nature03129},
	volume = {432},
	year = {2004}}

@article{Hegger2000,
  title = "{Pressure-Induced Superconductivity in Quasi-2D ${\mathrm{CeRhIn}}_{5}$}",
  author = {Hegger, H. and Petrovic, C. and Moshopoulou, E. G. and Hundley, M. F. and Sarrao, J. L. and Fisk, Z. and Thompson, J. D.},
  journal = {Phys. Rev. Lett.},
  volume = {84},
  issue = {21},
  pages = {4986--4989},
  numpages = {0},
  year = {2000},
  month = {May},
  publisher = {American Physical Society},
  doi = {10.1103/PhysRevLett.84.4986},
  url = {https://link.aps.org/doi/10.1103/PhysRevLett.84.4986}
}

@article{Millis1993,
  title = {Effect of a nonzero temperature on quantum critical points in itinerant fermion systems},
  author = {Millis, A. J.},
  journal = {Phys. Rev. B},
  volume = {48},
  issue = {10},
  pages = {7183--7196},
  numpages = {0},
  year = {1993},
  month = {Sep},
  publisher = {American Physical Society},
  doi = {10.1103/PhysRevB.48.7183},
  url = {https://link.aps.org/doi/10.1103/PhysRevB.48.7183}
}

@article{Scalapino1986,
  title = {$d$-wave pairing near a spin-density-wave instability},
  author = {Scalapino, D. J. and Loh, E. and Hirsch, J. E.},
  journal = {Phys. Rev. B},
  volume = {34},
  issue = {11},
  pages = {8190(R)--8192(R)},
  numpages = {0},
  year = {1986},
  month = {Dec},
  publisher = {American Physical Society},
  doi = {10.1103/PhysRevB.34.8190},
  url = {https://link.aps.org/doi/10.1103/PhysRevB.34.8190}
}

@article{Miyake1986,
  title = {Spin-fluctuation-mediated even-parity pairing in heavy-fermion superconductors},
  author = {Miyake, K. and Schmitt-Rink, S. and Varma, C. M.},
  journal = {Phys. Rev. B},
  volume = {34},
  issue = {9},
  pages = {6554(R)--6556(R)},
  numpages = {0},
  year = {1986},
  month = {Nov},
  publisher = {American Physical Society},
  doi = {10.1103/PhysRevB.34.6554},
  url = {https://link.aps.org/doi/10.1103/PhysRevB.34.6554}
}

@article{Hertz1976,
  title = {Quantum critical phenomena},
  author = {Hertz, John A.},
  journal = {Phys. Rev. B},
  volume = {14},
  issue = {3},
  pages = {1165--1184},
  numpages = {0},
  year = {1976},
  month = {Aug},
  publisher = {American Physical Society},
  doi = {10.1103/PhysRevB.14.1165},
  url = {https://link.aps.org/doi/10.1103/PhysRevB.14.1165}
}

@article{rende2025queries,
   title="{Are queries and keys always relevant? A case study on transformer wave functions}",
   volume={6},
   ISSN={2632-2153},
   url={http://dx.doi.org/10.1088/2632-2153/ada1a0},
   DOI={10.1088/2632-2153/ada1a0},
   number={1},
   journal={Machine Learning: Science and Technology},
   publisher={IOP Publishing},
   author={Rende, Riccardo and Loris Viteritti, Luciano},
   year={2025},
   month=Jan, pages={010501} }

@misc{viteritti2026bias,
      title="{Beyond Variational Bias: Resolving Intertwined Orders in the Hubbard Model}", 
      author={Luciano Loris Viteritti and Riccardo Rende and Christopher Roth and Anirvan Sengupta and Giuseppe Carleo and Antoine Georges},
      year={2026},
      eprint={2604.21978},
      archivePrefix={arXiv},
      primaryClass={cond-mat.str-el}
}

@article{sandvik1997prb,
  title = "{Finite-size scaling of the ground-state parameters of the two-dimensional Heisenberg model}",
  author = {Sandvik, Anders W.},
  journal = {Phys. Rev. B},
  volume = {56},
  issue = {18},
  pages = {11678--11690},
  numpages = {0},
  year = {1997},
  month = {Nov},
  publisher = {American Physical Society},
  doi = {10.1103/PhysRevB.56.11678},
  url = {https://link.aps.org/doi/10.1103/PhysRevB.56.11678}
}

@article{Lenz2017,
  title = {Variational cluster approach to superconductivity and magnetism in the Kondo lattice model},
  author = {Lenz, Benjamin and Gezzi, Riccardo and Manmana, Salvatore R.},
  journal = {Phys. Rev. B},
  volume = {96},
  issue = {15},
  pages = {155119},
  numpages = {24},
  year = {2017},
  month = {Oct},
  publisher = {American Physical Society},
  doi = {10.1103/PhysRevB.96.155119},
  url = {https://link.aps.org/doi/10.1103/PhysRevB.96.155119}
}

@article{Bodensiek2013,
  title = {Unconventional Superconductivity from Local Spin Fluctuations in the Kondo Lattice},
  author = {Bodensiek, Oliver and \ifmmode \check{Z}\else \v{Z}\fi{}itko, Rok and Vojta, Matthias and Jarrell, Mark and Pruschke, Thomas},
  journal = {Phys. Rev. Lett.},
  volume = {110},
  issue = {14},
  pages = {146406},
  numpages = {5},
  year = {2013},
  month = {Apr},
  publisher = {American Physical Society},
  doi = {10.1103/PhysRevLett.110.146406},
  url = {https://link.aps.org/doi/10.1103/PhysRevLett.110.146406}
}

@article{Hoshino2014,
  title = {Superconductivity of Composite Particles in a Two-Channel Kondo Lattice},
  author = {Hoshino, Shintaro and Kuramoto, Yoshio},
  journal = {Phys. Rev. Lett.},
  volume = {112},
  issue = {16},
  pages = {167204},
  numpages = {5},
  year = {2014},
  month = {Apr},
  publisher = {American Physical Society},
  doi = {10.1103/PhysRevLett.112.167204},
  url = {https://link.aps.org/doi/10.1103/PhysRevLett.112.167204}
}

@Article{LiuZhangYu2014,
title = {Pairing Symmetry of Heavy Fermion Superconductivity in the Two-Dimensional Kondo–Heisenberg Lattice Model},
journal = {Chin. Phys. Lett.},
volume = {31},
number = {8},
pages = {087102-087102},
year = {2014},
issn = {},
doi = {10.1088/0256-307X/31/8/087102},	
url = {http://cpl.iphy.ac.cn/en/article/doi/10.1088/0256-307X/31/8/087102},
author = {LIU Yu and ZHANG Guang-Ming and YU Lu}
}

@article{LiuHan2024,
  title = {Pair density wave and $\mathit{s}\ifmmode\pm\else\textpm\fi{}\mathit{id}$ superconductivity in a strongly coupled lightly doped Kondo insulator},
  author = {Liu, Fangze and Han, Zhaoyu},
  journal = {Phys. Rev. B},
  volume = {109},
  issue = {12},
  pages = {L121101},
  numpages = {5},
  year = {2024},
  month = {Mar},
  publisher = {American Physical Society},
  doi = {10.1103/PhysRevB.109.L121101},
  url = {https://link.aps.org/doi/10.1103/PhysRevB.109.L121101}
}

\newpage
\clearpage
\section*{END MATTER}
\section{Wave function and optimizer}

We parametrize the variational state using the Transformer-based neural-network quantum state for composite local Hilbert spaces introduced in Ref.~\cite{rende2026alm}. In the Kondo-Heisenberg Hamiltonian, each site carries a conduction electron and a localized spin-$1/2$ moment, so the local configuration $s_i=(n_{i\uparrow},n_{i\downarrow},S^z_i)$ spans a space of dimension $\mathcal{V}=8$, which we tokenize into integer labels $t_i$ and embed into vectors $x_i\in\mathbb{R}^d$. The sequence is processed by $n_l$ transformer layers with factored attention~\cite{viteritti2023prl,rende2025queries,rende2024prr,viteritti2025prb} and a distance-dependent spatial bias~\cite{viteritti2026approaching} that encodes the square-lattice geometry. The outputs $y_i\in\mathbb{R}^d$ define configuration-dependent backflow orbitals $\Phi_{i\sigma\alpha}=\sum_\beta y_{i\beta}W_{i\sigma\alpha\beta}$,
with $W\in\mathbb{R}^{N\times 2\times 2N\times d}$ a tensor of trainable
parameters; collecting $r=(i,\sigma)$, these form a matrix
$\phi_{r\alpha}\in\mathbb{R}^{2N\times 2N}$ (see Ref.~\cite{rende2026alm}).

To describe superconductive states efficiently, we antisymmetrize with a Pfaffian output layer~\cite{viteritti2026bias},
\begin{equation}
  \Psi_\theta(s)=\mathrm{Pf}\!\left[\,n\star\phi(s)\,A\,\phi(s)^{T}\!\star n\,\right],
  \label{eq:pfaffian}
\end{equation}
where $A\in\mathbb{R}^{2N\times 2N}$ is a trainable antisymmetric matrix and $n\star\phi$ selects the occupied rows. Unlike a Slater determinant, the Pfaffian encodes magnetic and superconducting channels simultaneously at the mean-field level, avoiding a bias against pairing given by determinant backflows~\cite{viteritti2026bias}.

In the superconducting region of the phase diagram, optimizations of \cref{eq:pfaffian} can remain trapped in the metastable stripe solutions discussed below. To avoid this, we seed the pairing channel by replacing the pairing matrix with $\phi(s)\,A\,\phi(s)^{T}+\Phi_d$, where $\Phi_d$ is a configuration-independent matrix. Its only nonzero elements connect opposite spins on nearest-neighbor bonds, $(\Phi_d)_{(\boldsymbol{r}\uparrow),(\boldsymbol{r}+\boldsymbol{\eta}\downarrow)}=\Delta_d\,h_{\boldsymbol{r},\boldsymbol{\eta}}$, with $\Delta_d$ a single trainable parameter and $h_{\boldsymbol{r},\boldsymbol{\eta}}=+1$ ($-1$) for $\boldsymbol{\eta}=\pm\hat{\boldsymbol{x}}$ ($\pm\hat{\boldsymbol{y}}$). This is the standard singlet $d_{x^2-y^2}$ pairing function, and the limit $\Delta_d\to0$ recovers \cref{eq:pfaffian} exactly.

The hyperparameters of the transformer architecture are chosen to be $n_l=4$ layers, $h=12$ heads, and $d=72$; see Ref.~\cite{viteritti2025prb} for their description. The state is optimized by Variational Monte Carlo with Stochastic Reconfiguration~\cite{sorella1998} using the linear algebra identity~\cite{rende2024stochastic,chen2024empowering} and the MARCH optimizer~\cite{gu2025}, over $20,000$ steps with $M=8192$ samples per iteration and a learning rate of $\tau=0.005$ annealed over time. The optimization is performed by fixing the total spin projection $S^z=0$~\cite{rende2026alm}. 

\section{Stripe order}
\begin{figure}[t]
    \begin{center}
\centerline{\includegraphics[width=0.8\columnwidth]{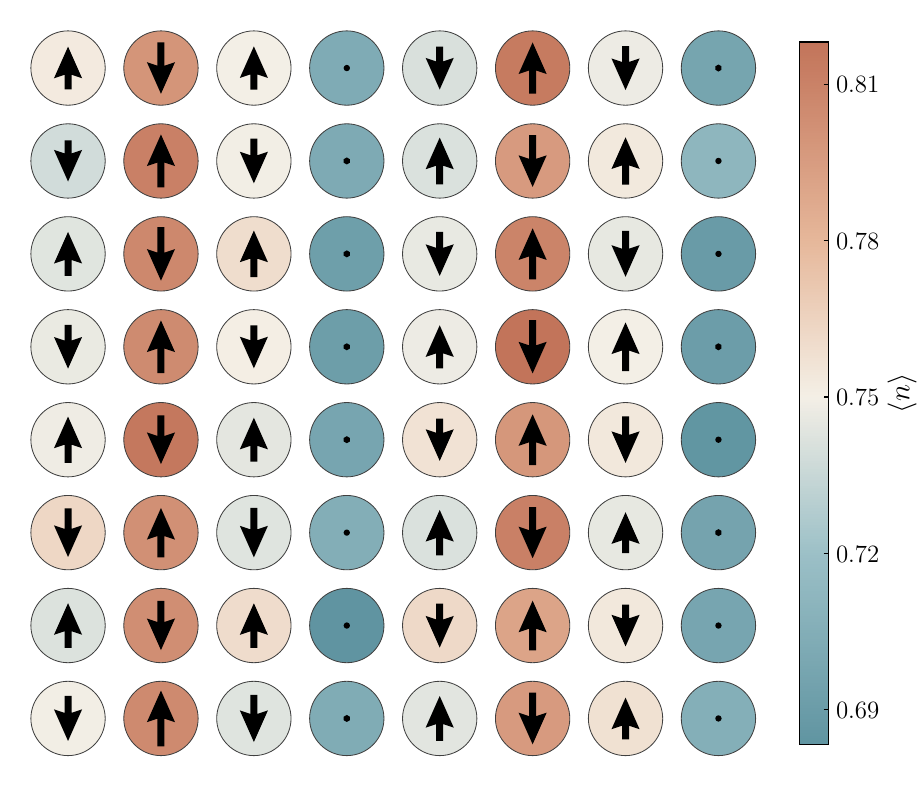}}
    \caption{Local density $\langle\hat{n}_i\rangle$ (colour) and moment orientation (arrows) for the stripe state at $J=0$ and $J_K=1.4$.} \label{fig:stripe}
    \end{center}
\end{figure}

At $J=0$, the conduction electrons develop stripe order, shown in \cref{fig:stripe} for $J_K=1.4$. The snapshot of the local density $\langle\hat{n}_i\rangle$ and the moment orientation displays a unidirectional modulation: lines of reduced density separate magnetic domains in which the order reverses phase, so that the density minima coincide with antiphase domain walls. This is the same stripe order reported for the $t$--$t'$ Hubbard model at $t'=-0.2$, $U=8$, and hole doping $1/8$~\cite{gu2025,viteritti2026bias}. This stripe state is not confined to $J=0$: for finite $J$ it is also obtained as a metastable solution to which some optimization seeds converge, but its variational energy is always higher than that of the antiferromagnetic-plus-superconducting solution, which we therefore take as the ground state. We do not study the extent of the stripe region in the phase diagram in this work.

\section{Mean-field details}
The localized spins are represented using fermionic spinons 

\begin{equation}
\boldsymbol S_i
=
\frac12
\sum_{\alpha\beta}
f^\dagger_{i\alpha}
\boldsymbol\tau_{\alpha\beta}
f_{i\beta},
\label{eq:parton}
\end{equation}
subject to the single-occupancy constraint $\sum_\sigma f^\dagger_{i\sigma}f_{i\sigma}=1$. At the mean-field level, the dominant competing channels are described by the spinon hopping amplitude $\chi_{ij}=\langle f^\dagger_{i\sigma}f_{j\sigma}\rangle$, the Kondo hybridization $V_i=\langle c^\dagger_{i\sigma}f_{i\sigma}\rangle$, and the staggered antiferromagnetic order parameters
\begin{equation}
\langle S_i^z\rangle
=
m_s e^{iQ\cdot R_i},
\quad
\langle s_i^z\rangle
=
m_e e^{iQ\cdot R_i}   , 
\end{equation} 
with ordering vector $Q=(\pi,\pi)$. The saddle-point parameters $(\chi,V,\lambda,m_s,m_e)$ are obtained by solving self-consistent equations.

Assuming a uniform saddle point, $\chi_{ij}=\chi$, $V_i=V$, and $\lambda_i=\lambda$, the mean-field Hamiltonian can be written in the reduced Brillouin zone as
\begin{equation}
H_{\rm MF}
=
\sum_{k\in{\rm RBZ},\sigma}
\Psi_{k\sigma}^\dagger
\mathcal H_{k\sigma}
\Psi_{k\sigma},
\end{equation}
where $\Psi_{k\sigma}^T=(c_{k\sigma},c_{k+Q,\sigma},f_{k\sigma},f_{k+Q,\sigma})$ and
\begin{equation}
\mathcal H_{k\sigma}
=
\begin{pmatrix}
\epsilon_k^c & \sigma \Delta_c & \phi & 0 \\
\sigma \Delta_c & \epsilon_{k+Q}^c & 0 & \phi \\
\phi & 0 & \epsilon_k^f & \sigma \Delta_f \\
0 & \phi & \sigma \Delta_f & \epsilon_{k+Q}^f
\end{pmatrix}.
\label{eq:Hmf}
\end{equation}

The dispersions of the two layers are given by
\begin{equation}
\epsilon_k^c=-t\Gamma(k)-\mu, \quad
\epsilon_k^f
=
\lambda
-
\frac{J}{2}\chi\Gamma(k),
\end{equation}
with lattice form factor $\Gamma(k)=2(\cos k_x+\cos k_y)$. The parameter $\phi=-J_KV/2$ describes Kondo hybridization, while
\begin{equation}
\Delta_c
=
\frac{J_K}{2}m_s,
\qquad
\Delta_f
=
-
\frac{zJ}{2}m_s
+
\frac{J_K}{2}m_e
\end{equation}
describe the staggered magnetic order.

The first two self-consistency equations are given by
\begin{align}
1-p &= \frac{1}{N} \sum_{k,\sigma}
\langle c^\dagger_{k\sigma} c_{k\sigma} \rangle, \label{eq:constraint_c} \\
1 &= \frac{1}{N} \sum_{k,\sigma}
\langle f^\dagger_{k\sigma} f_{k\sigma} \rangle, \label{eq:constraint_f}
\end{align}
where $N$ is the number of discretized points which we take to be $N=120\times 120$ for most of the numerical simulations. The equations fix the density of conduction electrons and satisfy the spinon single-occupancy constraint. The spinon hopping amplitude and hybridization are

\begin{figure}[t]
    \begin{center}
\centerline{\includegraphics[width=0.8\columnwidth]{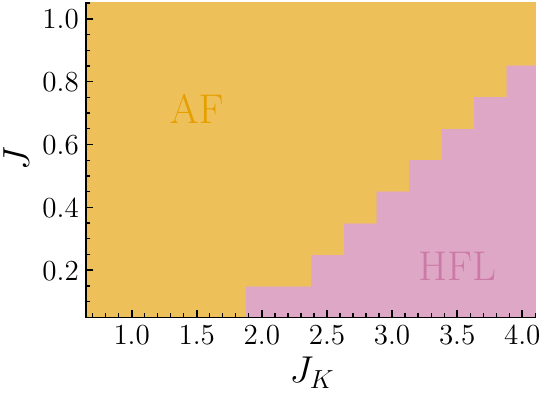}}
        \caption{Mean-field phase diagram of the two-dimensional Kondo-Heisenberg model in the $(J_K, J)$ plane. Two phases are identified: an antiferromagnetic N\'eel phase (AF) and a heavy Fermi liquid (HFL). }
        \label{fig:phase_diagram_MF}
    \end{center}
\end{figure}

\begin{align}
\chi &= \frac{1}{N z} \sum_{k,\sigma}
\Gamma(k)\,
\langle f^\dagger_{k\sigma} f_{k\sigma} \rangle, \\
V &= \frac{1}{N} \sum_{k,\sigma}
\langle c^\dagger_{k\sigma} f_{k\sigma} \rangle, 
\end{align}
where $z=4$ is the coordination number of the square lattice and $\Gamma(k)=2(\cos k_x+\cos k_y)$.
Finally, the staggered magnetizations in both layers are determined by hybridization between $k$ and $k+Q$ states
\begin{equation}
m_s
=
\frac{1}{2N}
\sum_{k,\sigma}
\sigma\,
\langle f^\dagger_{k+Q,\sigma} f_{k\sigma} \rangle,
\,
m_e
=
\frac{1}{2N}
\sum_{k,\sigma}
\sigma\,
\langle c^\dagger_{k+Q,\sigma} c_{k\sigma} \rangle.
\end{equation}
All the averages are computed by diagonalizing the mean-field Hamiltonian in Eq.~(\ref{eq:Hmf}) and taking thermal averages at $T=0.01$ for all calculations.

The free energy is obtained from the quasiparticle spectrum of the Hamiltonian:
\begin{equation}
F
=
E_0
-
T \sum_{k, n}
\ln\left(1 + e^{-E_{kn}/T}\right),
\end{equation}

where \(E_{kn}\) are the eigenvalues of \(\mathcal H_{k\sigma}\) and $T$ is the temperature. The constant term is given by 

\begin{equation}
E_0/N
=
- \lambda
+
\frac{ z J}{4}\chi^2
+
\frac{J_K}{2}V^2
+
 \frac{ zJ}{2} m_s^2
-
 J_K m_s m_e+\mu(1-p).
\end{equation}
We note that self-consistency equations can be alternatively derived from the free energy, since the solution corresponds to the local minima $\partial F/\partial x_i=0$.
For a given set of parameters \((J,J_K,p,T)\), the self-consistent solution is obtained by solving the above equations simultaneously. When multiple solutions exist, the physical state is identified as the one with the lowest free energy.

At small $J_K$ we find the AF solution to have the lowest energy, while at large $J_K$ the heavy Fermi liquid solution dominates, see Fig.~\ref{fig:phase_diagram_MF}. The phase boundaries match very well with the full numerical solution, cf. Fig.~\ref{fig:phase_diagram}. The transition between the two phases is of first order.

\appendix

\newpage

\widetext

\section{FL$^*$ and SC mean-field}

While the previous mean-field ansatz naturally explains the transition between the antiferromagnetic state and the heavy Fermi liquid, it does not capture the superconducting tendencies observed near the quantum critical region. To describe this competing phase, we consider an alternative parton construction with an emergent $SU(2)$ gauge symmetry, in which the local moments form a quantum spin liquid. The localized moments are represented by fermionic spinons, subject to the local constraint, see Eq.~\ref{eq:parton}. 
The spin-liquid mean-field ansatz in the spin layer is written as
\begin{equation}
\hat H_f^{\rm MF}
=
\frac{J}{4}
\sum_{\langle ij\rangle}
\left(
F_i^\dagger U_{ij} F_j
+
\mathrm{h.c.}
\right),
\end{equation}
, where we introduce the Nambu spinors
\(
F_i=(f_{i\uparrow},f^\dagger_{i\downarrow})^T
\)
and
\(
C_i=(c_{i\uparrow},c^\dagger_{i\downarrow})^T,
\) with
\begin{equation}
U_{ij}
=
\begin{pmatrix}
\chi & e_{ij}\Delta_f\\
e_{ij}\Delta_f & -\chi
\end{pmatrix},
\qquad
e_{ij}
=
\begin{cases}
+1, & ij\parallel x,\\
-1, & ij\parallel y .
\end{cases}
\end{equation}
The sign structure of $e_{ij}$ corresponds to $d$-wave spinon pairing.
We also introduce a bosonic chargon matrix that couples the two layers
\begin{equation}
B_i
=
\begin{pmatrix}
B_{1,i} & B_{2,i}\\
B^*_{2,i} & -B^*_{1,i}
\end{pmatrix}.
\end{equation}
At the mean-field level, the Kondo term becomes
\begin{equation}
\hat H_K^{\rm MF}
=
\frac{J_K}{4}
\sum_i
\left(
F_i^\dagger B_i C_i
+
\mathrm{h.c.}
\right)
 .
\end{equation}

Assuming a uniform saddle point, the momentum-space mean-field Hamiltonian takes the Bogoliubov-de Gennes form
\begin{equation}
\hat H_{\rm MF}
=
E_0+
\sum_k
\Psi_k^\dagger
\mathcal H_k
\Psi_k ,
\end{equation}
where the basis vector
\(
\Psi_k=
(c_{k\uparrow},
c^\dagger_{-k\downarrow},
f_{k\uparrow},
f^\dagger_{-k\downarrow})^T
\)
and the corresponding mean-field matrix is
\begin{equation}
\mathcal H_k
=
\begin{pmatrix}
\xi_k^c & 0 & \Phi & 0\\
0 & -\xi_k^c & 0 & -\Phi\\
\Phi & 0 & \xi_k^f & \Delta_k^f\\
0 & -\Phi & \Delta_k^f & -\xi_k^f
\end{pmatrix}.
\label{eq:flstar_bdg_hamiltonian}
\end{equation}
Here
\(
\xi_k^f=\lambda-\frac{J \chi}{2}\Gamma(k),
\)
\(
\Delta_k^f=\frac{J\Delta_f}{2}\Gamma_d(k),
\)
with the lattice form factors
\(
\Gamma(k)=2(\cos k_x+\cos k_y)
\)
and
\(
\Gamma_d(k)=2(\cos k_x-\cos k_y).
\)
Note that $\Gamma_d(k)$ automatically incorporates $d$-wave symmetry since the order parameter vanishes on the diagonal.
We further assume that $B_{2,i}=0$ and thus
\(
\Phi=\frac{J_K}{4}B_1
\).

So far, we have omitted the presence of antiferromagnetic order. It could be naturally incorporated in the mean-field theory by extending the basis to 
\(
\Psi^{\mathrm{AF}}_k=
(c_{k\uparrow},
c^\dagger_{-k\downarrow},
f_{k\uparrow},
f^\dagger_{-k\downarrow},c_{k+Q\uparrow},
c^\dagger_{-k-Q\downarrow},
f_{k+Q\uparrow},
f^\dagger_{-k-Q\downarrow})^T
\), where $Q=(\pi,\pi)$ is the ordering vector.
In block form, the Hamiltonian becomes
\begin{equation}
\mathcal H^{\mathrm{AF}}_k=
\begin{pmatrix}
\mathcal H_k & \mathcal V \\
\mathcal V^\dagger & \mathcal H_{k+Q}
\end{pmatrix},
\end{equation}

where the staggered field matrix is given by
\begin{equation}
\mathcal V=
\begin{pmatrix}
\Delta^{\mathrm{AF}}_c & 0 & 0 & 0\\
0 & \Delta^{\mathrm{AF}}_c & 0 & 0\\
0 & 0 & \Delta^{\mathrm{AF}}_f & 0\\
0 & 0 & 0 & \Delta^{\mathrm{AF}}_f
\end{pmatrix}.
\end{equation}
As previously, the staggered fields are expressed in terms of magnetizations:
\begin{equation}
\Delta^{\mathrm{AF}}_c=\frac{J_K}{2}m_s,
\qquad
\Delta_f^{ \mathrm{AF}}= -\frac{z J}{2}m_s+\frac{J_K}{2}m_e.
\end{equation}
After introducing the mean-field Hamiltonian, we turn our attention to writing the self-consistency equations.
The spinon constraint and the conduction electron density are described by the same Eq.~(\ref{eq:constraint_c},\ref{eq:constraint_f}).

The spinon hopping and pairing equations are

\begin{align}
\chi
=
\frac{1}{N z}
\sum_{k,\sigma}
\Gamma(k)
\langle
\hat f^\dagger_{k\sigma}\hat f_{k\sigma}
\rangle ,\\
\Delta_f
=
\frac{1}{N z}
\sum_k
\Gamma_d(k)
\langle
\hat f_{-k\downarrow}
\hat f_{k\uparrow}
\rangle ,
\end{align}

The chargon equation is

\begin{equation}
B_1
=
\frac1N
\sum_k
\left[
\langle
\hat f^\dagger_{k\uparrow}
\hat c_{k\uparrow}
\rangle
+
\langle
\hat f_{-k\downarrow}
\hat c^\dagger_{-k\downarrow}
\rangle
\right].
\end{equation}

\begin{figure}[t]
    \begin{center}
\centerline{\includegraphics[width=0.5\columnwidth]{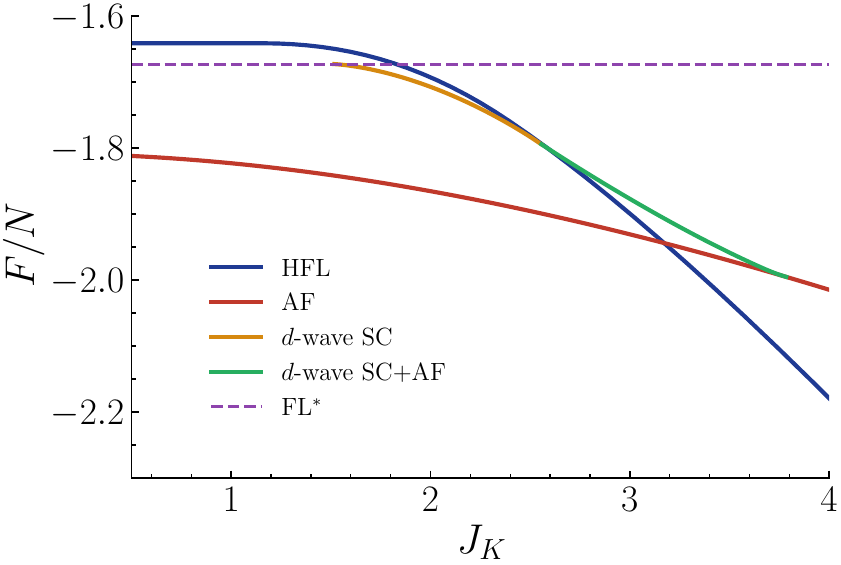}}
    \caption{Free energy of different mean-field ansatzes as a function of $J_K$ at fixed $J=0.5$ and temperature $T=0.01$.} \label{fig:free_energy}
    \end{center}
\end{figure}

Finally, the antiferromagnetic order parameters satisfy
\begin{equation}
m_s=
\frac{1}{2N}
\sum_{k,\sigma}
\sigma
\langle f^\dagger_{k+Q,\sigma}f_{k\sigma}\rangle,
\qquad
m_e=
\frac{1}{2N}
\sum_{k,\sigma}
\sigma
\langle c^\dagger_{k+Q,\sigma}c_{k\sigma}\rangle .
\end{equation}

\begin{equation}
F=E_0+
\sum_{k\in\mathrm{RBZ}}
\left(
\xi_k^c+
\xi_{k+Q}^c+
\xi_k^f+
\xi_{k+Q}^f
\right)
-T\sum_{k\in\mathrm{RBZ}}
\sum_{n=1}^{8}
\ln\left[1+e^{-E_{kn}/T}\right].
\end{equation}
where \(E_{kn}\) are the eigenvalues of \(\mathcal H_{k}\) and the summation over $k$ happens in the reduced Brillouin zone. The Second term corresponds to the vacuum energy shift, typical in Bogoliubov-de Gennes formalism. The constant term is
\begin{equation}
E_0/N=
-\lambda
+
z\frac{J}{4}
\left(\chi^2+\Delta_f^2\right)
+\frac{J_K}{8}B_1^2
-z J m_s^2
-J_Km_sm_e
+\mu(1-p).
\end{equation}

At small $J_K$, we find a decoupled phase with $B_1=0$ and no magnetic order, while the spinon sector forms a paired Dirac spin liquid with
\(
\Delta_f=\chi .
\)
Such a state corresponds to a fractionalized Fermi liquid, denoted FL$^*$. In this phase, the conduction electrons form a small Fermi surface while the local moments remain fractionalized into neutral spinons.

As $J_K$ increases, the chargon field condenses and induces superconducting order. The resulting SC phase has $d$-wave pairing symmetry, consistent with NQS results.
At larger $J_K$, the magnetizations become nonzero, and the superconducting state coexists with antiferromagnetic order until it transitions to a heavy Fermi liquid, see Fig.~\ref{fig:free_energy}. 

We note that within the present mean-field treatment the N\'eel antiferromagnetic state always has a lower free energy than either the FL$^*$ or superconducting solutions. This result should be interpreted with caution, since mean-field approximations are known to overestimate the stability of long-range ordered phases and underestimate the role of quantum fluctuations. Overall, it is encouraging that the same mean-field framework naturally produces a superconducting solution in close proximity to the antiferromagnetic state, consistent with NQS simulations.

\section{Structure factor}
In this appendix, we compute equal-time structure factor in the heavy Fermi liquid regime and compare it to the NQS simulations. We focus on the regime where the antiferromagnetic order is absent, and the mean-field Hamiltonian is given by:
\begin{equation}
\hat H_{\rm MF}
=
E_0+
\sum_{k,\sigma}
\begin{pmatrix}
\hat c^\dagger_{k\sigma} &
\hat f^\dagger_{k\sigma}
\end{pmatrix}
\begin{pmatrix}
\epsilon_k^c & \phi\\
\phi & \epsilon_k^f
\end{pmatrix}
\begin{pmatrix}
\hat c_{k\sigma}\\
\hat f_{k\sigma}
\end{pmatrix}.
\end{equation}

As previously, the dispersions in the electron and spin layers are given by
\(
\epsilon_k^c=-t\Gamma(k)-\mu, \quad
\epsilon_k^f
=
\lambda
-
\frac{J}{2}\chi\Gamma(k),
\)
. The lattice form factor $\Gamma(k)=2(\cos k_x+\cos k_y)$ and the parameter $\phi=-J_KV/2$ describes Kondo hybridization.
The mean-field Hamiltonian is diagonalizable with the dispersion
\begin{equation}
E_k^\pm
=
\frac{\epsilon_k^c+\epsilon_k^f}{2}
\pm
\sqrt{
\left(
\frac{\epsilon_k^c-\epsilon_k^f}{2}
\right)^2
+
\phi^2
}.
\end{equation}

The equal time Green's functions are expressed in terms of coherence factors as

\begin{equation}
G_{cc}(k) = \langle c^\dagger_{k\sigma} c_{k\sigma} \rangle
=
u_k^2 n_k^+
+
v_k^2 n_k^- ,
\end{equation}

\begin{equation}
G_{ff}(k) = \langle f^\dagger_{k\sigma} f_{k\sigma} \rangle
=
v_k^2 n_k^+
+
u_k^2 n_k^- ,
\end{equation}

\begin{equation}
G_{cf}(k) = \langle c^\dagger_{k\sigma} f_{k\sigma} \rangle
=
u_kv_k
\left(
n_k^+
-
n_k^-
\right).
\end{equation}
, where 
\(
n_k^\pm
=
n_F(E_k^\pm), 
\) and $n_F(\epsilon)=1/(1+e^{\epsilon/T})$ is the Fermi-Dirac distribution. The coherence factors are

\begin{equation}
u_k^2=
\frac12
\left[
1+
\frac{\epsilon_k^c-\epsilon_k^f}
{\sqrt{(\epsilon_k^c-\epsilon_k^f)^2+4\phi^2}}
\right], \quad
v_k^2=
\frac12
\left[
1-
\frac{\epsilon_k^c-\epsilon_k^f}
{\sqrt{(\epsilon_k^c-\epsilon_k^f)^2+4\phi^2}}
\right].
\end{equation}

The structure factors can be represented in terms of equal time Green's functions as

\begin{equation}
S_{e}({\bf q})
=
\frac{3}{2N}
\sum_k
G_{cc}(k+q)
\Bigl[
1-G_{cc}(k)
\Bigr],
\end{equation}

\begin{equation}
S({\bf q})
=
\frac{3}{2N}
\sum_k
G_{ff}(k+q)
\Bigl[
1-G_{ff}(k)
\Bigr],
\end{equation}

\begin{figure}[]
    \begin{center}
\centerline{\includegraphics[width=1\columnwidth]{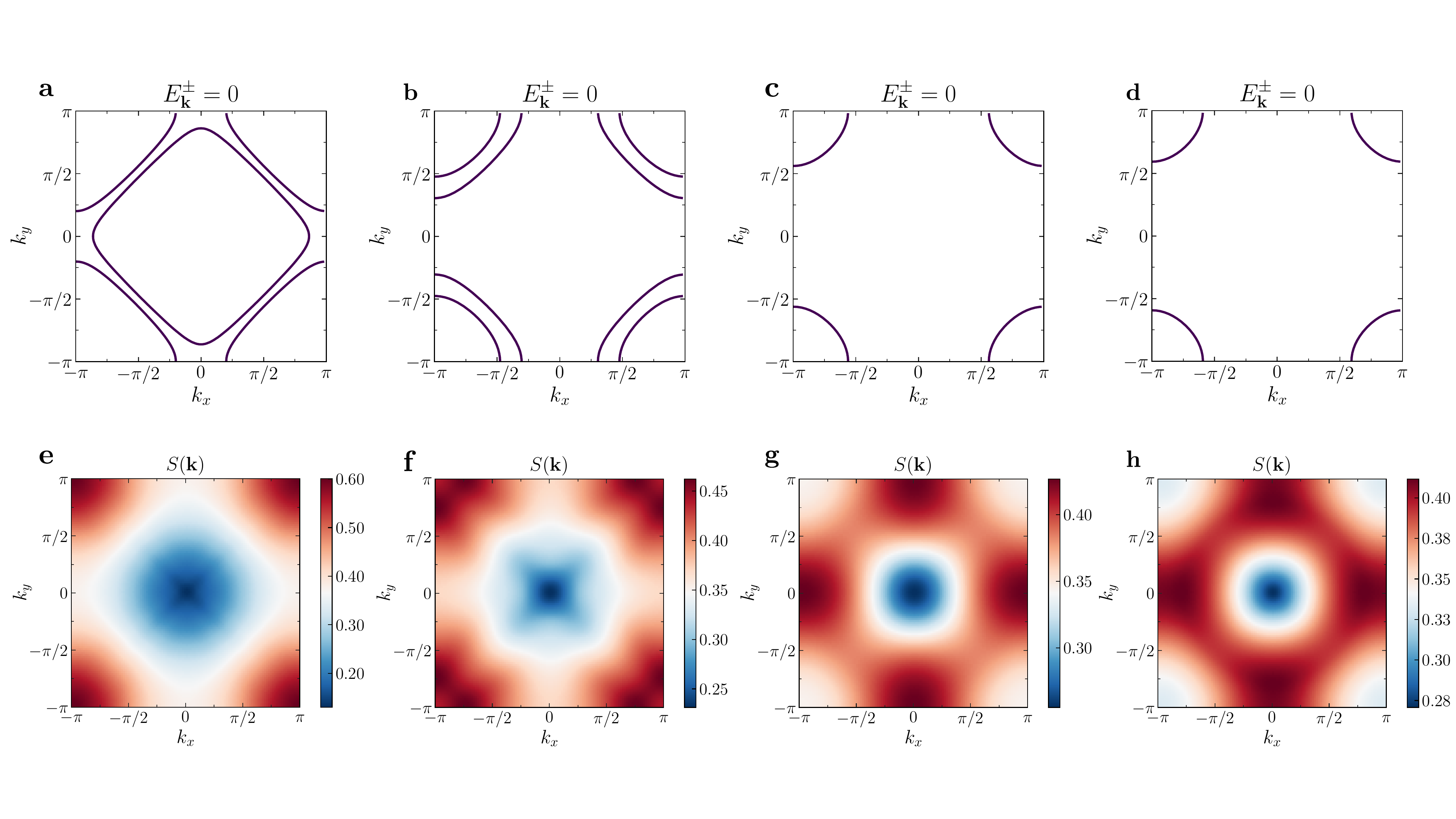}}
        \caption{
        Top row ($\boldsymbol{a},\boldsymbol{b},\boldsymbol{c},\boldsymbol{d}$): the hybridized bands at zero energy.
        Bottom row($\boldsymbol{e},\boldsymbol{f},\boldsymbol{g},\boldsymbol{h}$): mean-field static structure factors of localized spins,
$S(\boldsymbol k)$.
From left to right, the panels correspond to Kondo hybridizations
$\phi=0.23$,$0.57$, $0.82$, and $1.54$, respectively. }
        \label{fig:structure_factor_MF}
    \end{center}
\end{figure}

Fig.~\ref{fig:structure_factor_MF} shows the evolution of the structure factor within the heavy Fermi liquid phase as we increase Kondo hybridization. 
At small $\phi$, the structure factor is peaked at  $Q=(\pi,\pi)$, which is a consequence of nesting in the spinon Fermi surface. The hole-like regions are centered across the diagonal, see Fig.~\ref{fig:structure_factor_MF} (a,b).
This regime is close to the critical region, see Fig.~\ref{fig:Sk_nk} (c,g), where the structure factor is similarly peaked at $Q=(\pi,\pi)$ despite the lack of the long-range order and the Fermi surface remains small.

At $\phi \approx0.74$, there is a Lifshitz transition, the hybridized bands are reconstructed, and a hole pocket appears at $Q=(\pi,\pi)$. The structure factor becomes peaked at $2k_F$ which is the result of the Kohn anomaly. 

\section{Finite size mean-field}

\begin{figure}[b]
\begin{minipage}[h]{0.6\linewidth}
  \center{\includegraphics[width=1\linewidth]{
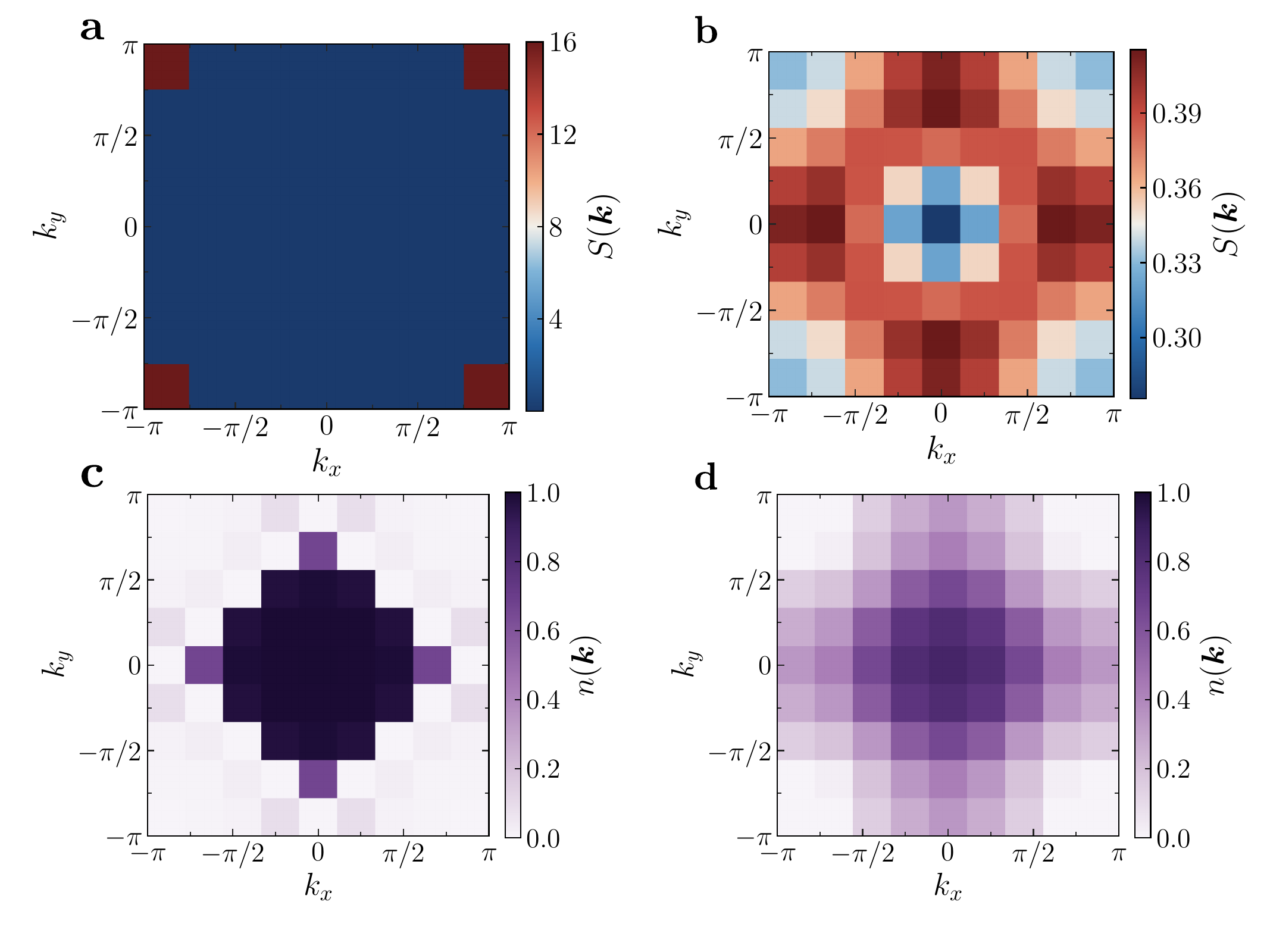}}
\end{minipage}
\caption{Mean field local-moment structure factor $S(\boldsymbol{k})$ (top row) and conduction-electron momentum distribution $n(\boldsymbol{k})$ (bottom row) for $8 \times 8$ lattice, at fixed $J = 0.5$ and three representative values of the Kondo coupling: $J_K = 2.0$ (left, panels $\boldsymbol{a},\boldsymbol{c}$) and $J_K = 4.0$ (right, panels $\boldsymbol{b},\boldsymbol{d}$).}
\label{fig:mean_field_nk_finite}
\end{figure}

To further compare the NQS simulations with the mean-field results, we compute the mean-field static spin structure factor $S(\mathbf{k})$ and the conduction-electron momentum distribution $n(\mathbf{k})$ on a finite $N=8 \times 8$ lattice, as shown in Fig.~\ref{fig:mean_field_nk_finite}. The conduction-electron momentum distribution exhibits a transition from a small to a large Fermi surface and agrees quantitatively with the NQS results shown in Fig.~\ref{fig:Sk_nk}(f,h). The spin structure factor in the antiferromagnetic phase displays a pronounced peak at $\mathbf{Q}=(\pi,\pi)$. Its magnitude is larger than that obtained in the corresponding NQS simulations, indicating that the mean-field theory overestimates the strength of the antiferromagnetic order. At $J_K=4$, the structure factor develops a peak at $2k_F$, in agreement with Fig.~\ref{fig:Sk_nk}(d).

\end{document}